\documentclass[aps,prd,twocolumn,superscriptaddress]{revtex4}
\usepackage{epsfig,epsf}
\usepackage{amsmath}
\usepackage{amsthm}
\usepackage{amsfonts}
\usepackage{amssymb}
\usepackage{dsfont}
\usepackage{multirow}
\usepackage{appendix}
\usepackage{slashed}
\usepackage[active]{srcltx}
\usepackage{psfrag}

\begin{document}

\title{On the structures of new scalar resonances $T_{cs0}^{a}(2900)^{++}$
and $T_{cs0}^{a}(2900)^{0}$ }
\date{\today}
\author{S.~S.~Agaev}
\affiliation{Institute for Physical Problems, Baku State University, Az--1148 Baku,
Azerbaijan}
\author{K.~Azizi}
\affiliation{Department of Physics, University of Tehran, North Karegar Avenue, Tehran
14395-547, Iran}
\affiliation{Department of Physics, Do\v{g}u\c{s} University, Dudullu-\"{U}mraniye, 34775
Istanbul, Turkiye}
\author{H.~Sundu}
\affiliation{Department of Physics, Kocaeli University, 41380 Izmit, Turkiye}
\affiliation{Department of Physics Engineering, Istanbul Medeniyet University, 34700
Istanbul, Turkiye}

\begin{abstract}
We investigate properties of the new scalar resonances $%
T_{cs0}^{a}(2900)^{++}$ and $T_{cs0}^{a}(2900)^{0}$, which were recently
reported by the LHCb collaboration. These states were observed as resonant
structures in $D_{s}^{+}\pi^{+}$ and $D_{s}^{+}\pi^{-}$ invariant mass
distributions in $B^{+}$ meson decays. We argue that $T_{cs0}^{a}(2900)^{++}
$ and $T_{cs0}^{a}(2900)^{0}$ may be modeled as molecules $\mathcal{M}%
^{++}=D_{s}^{\ast+}\rho^{+}$ and $\mathcal{M}^{0}=D_{s}^{\ast+}\rho^{-} $ of
conventional vector mesons, respectively. The mass $m$ and current coupling $%
f $ of the molecule $\mathcal{M}^{++}$ are calculated using two-point sum
rule method. The sum rule analysis is performed by taking into account
vacuum condensates up to dimension $8$. The obtained result for the mass, $%
m=(2917 \pm 135)~\mathrm{MeV}$, permits us to consider the molecule $%
\mathcal{M}^{++}$ as one of possible models of the resonance $%
T_{cs0}^{a}(2900)^{++}$. Because the second structure $T_{cs0}^{a}(2900)^{0}$
is isospin partner of the doubly charged state, it should have the mass
close to $m$.
\end{abstract}

\maketitle


\section{Introduction}

\label{sec:Intro}

A few months ago the LHCb collaboration reported about observation of three
new exotic hadrons labeled as $P^{\psi s}(4338)^{0}$, $T_{cs0}^{a}(2900)^{0}$
and $T_{cs0}^{a}(2900)^{++}$, respectively \cite{LHCb:2022xob,LHCb:2022bkt}.
First of them is presumably a pentaquark discovered with a high significance
in the $J/\psi \Lambda $ invariant mass distribution in the decay $%
B^{-}\rightarrow J/\psi \Lambda p$. Remaining two resonances are four-quark
mesons $T_{cs0}^{a}(2900)^{0}$ and $T_{cs0}^{a}(2900)^{++}$ (hereafter $%
T_{cs0}^{a0}$ and $T_{cs0}^{a++}$, respectively) fixed in the processes $%
B^{0}\rightarrow D^{0}D_{s}^{+}\pi ^{-}$ and $B^{+}\rightarrow
D^{-}D_{s}^{+}\pi ^{+}$. They are scalar particles and were seen in the mass
distributions of the mesons $D_{s}^{+}\pi ^{-}$ and $D_{s}^{+}\pi ^{+}$. The
LHCb measured the masses and widths of the exotic mesons $T_{cs0}^{a0}$
\begin{eqnarray}
m_{1\mathrm{exp}} &=&(2892\pm 14\pm 15)~\mathrm{MeV},  \notag \\
\Gamma _{1\mathrm{exp}} &=&(119\pm 26\pm 13)~\mathrm{MeV},  \label{eq:DataT0}
\end{eqnarray}%
and $T_{cs0}^{a++}$
\begin{eqnarray}
m_{2\mathrm{exp}} &=&(2923\pm 17\pm 20)~\mathrm{MeV},  \notag \\
\Gamma _{2\mathrm{exp}} &=&(137\pm 32\pm 17)~\mathrm{MeV}.  \label{eq:DtaT++}
\end{eqnarray}%
The collaboration also provided the following data for these structures:
\begin{eqnarray}
m_{\mathrm{exp}} &=&(2908\pm 11\pm 20)~\mathrm{MeV},\   \notag \\
\Gamma _{\mathrm{exp}} &=&(136\pm 23\pm 11)~\mathrm{MeV},  \label{eq:Data0}
\end{eqnarray}%
supposing that they are isospin partners and share these parameters.

Because the new scalar resonances were fixed in the $D_{s}^{+}\pi ^{-}$ and $%
D_{s}^{+}\pi ^{+}$ mass distributions, the processes $T_{cs0}^{a0}%
\rightarrow D_{s}^{+}\pi ^{-}$ and $T_{cs0}^{a++}\rightarrow D_{s}^{+}\pi
^{+}$ are their main decay channels. Then, it is clear that they are built
of quarks $cd\overline{s}\overline{u}$ and $cu\overline{s}\overline{d}$,
provided the four-quark picture can be applied to model these states. They
are also fully open flavor tetraquarks, and hence belong to a family of
exotic mesons established by the structures $X_{0}(2900)$ and $X_{1}(2900)$.
In fact, in Refs.\ \cite{LHCb:2020A,LHCb:2020} the LHCb informed on scalar $%
X_{0}(2900)$ and vector $X_{1}(2900)$ resonances (in what follows $X_{0}$
and $X_{1}$) in the invariant mass distribution of $D^{-}K^{+}$ mesons in
the channel $B^{+}\rightarrow D^{+}D^{-}K^{+}$. This means that the
resonances $X_{0}$ and $X_{1}$ decay to mesons $D^{-}K^{+}$, and in the
four-quark model are composed of four different valence quarks $ud\overline{s%
}\overline{c}$. This fact placed $X_{0}$ and $X_{1}$ to special position in $%
XYZ$ family of exotic mesons, because they were first evidences for the
fully open flavor tetraquarks.

The discovery of the resonances $X_{0}$ and $X_{1}$ triggered interesting
theoretical investigations aimed to understand their nature, calculate their
masses, and estimate widths of these states \cite%
{Karliner:2020vsi,Wang:2020xyc,Chen:2020aos,Liu:2020nil,Molina:2020hde,Hu:2020mxp,He:2020jna, Lu:2020qmp,Zhang:2020oze,Huang:2020ptc,Xue:2020vtq, Yang:2021izl,Wu:2020job,Abreu:2020ony,Wang:2020prk,Xiao:2020ltm,Dong:2020rgs,Burns:2020xne,Bondar:2020eoa, Chen:2020eyu,Albuquerque:2020ugi}%
. Naturally, authors of numerous publications made different assumptions
about internal structures of $X_{0}$ and $X_{1}$, and invoked various models
and calculational schemes to evaluate their parameters. For instance, in
Refs.\ \cite{Karliner:2020vsi,Wang:2020xyc} $X_{0}$ was treated as a scalar
diquark-antidiquark state $[sc][\overline{u}\overline{d}]$. Contrary, in
Ref.\ \cite{Chen:2020aos} $X_{0}$ was assigned as the $S$-wave hadronic
molecule $D^{\ast -}K^{\ast +}$, whereas for $X_{1}$ the authors adopted a
diquark-antidiquark model.

We studied the resonance $X_{0}$ by considering it as a molecule $\overline{D%
}^{\ast 0}K^{\ast 0}$ and evaluated its mass and width \cite{Agaev:2020nrc}.
Obtained results allowed us to confirm the molecule nature of $X_{0}$. We
explored in Ref.\ \cite{Agaev:2021knl} also the resonance $X_{1}$ and
interpreted it as a vector diquark-antidiqurk state $X_{\mathrm{V}}=[ud][%
\overline{c}\overline{s}]$. The discovery of $X_{0}$ and $X_{1}$ by LHCb
made the diquark-antidiquarks $[ud][\overline{c}\overline{s}]$ objects of
intensive investigations. Indeed, the vector state $X_{\mathrm{V}}$ from
this group of particles presumably was seen in experiment as the resonance $%
X_{1}$. The masses and full widths of the ground-state and radially excited
scalar particles $X_{0}^{(\prime )}=[ud][\overline{c}\overline{s}]$ were
computed in Ref.\ \cite{Agaev:2022eeh}. Spectroscopic parameters and widths
of the axial-vector and pseudoscalar tetraquarks $X_{\mathrm{AV}}$ and $X_{%
\mathrm{PS}}$ with the same content were calculated \cite{Sundu:2022kyd}, as
well.

It is worth to emphasize that the fully open flavor exotic mesons were
already in agenda of researchers. Thus, the scalar diquark-antidiquark state
$X_{c}=[su][\overline{c}\overline{d}]$ was investigated in Ref.\ \cite%
{Agaev:2016lkl}, in which its mass and full width were computed in the
framework of QCD sum rule method using the $C\gamma _{5}\otimes \gamma _{5}C$
and $C\gamma _{\mu }\otimes \gamma ^{\mu }C$ structures and, accordingly,
the scalar-scalar and axial-axial interpolating currents. The scalar,
pseudoscalar and axial-vector diquark-antidiquark states $[sd][\overline{u}%
\overline{c}]$, which carry two units of electric charge $-2|e|$ were
studied in Ref. \cite{Agaev:2017oay}. The particles $Z^{++}$ with the
content $[cu][\overline{s}\overline{d}]$ are positively charged counterparts
of these states and have the same masses and decay widths. Parameters of the
vector tetraquark $Z_{\mathrm{V}}^{++}$ became available recently \cite%
{Agaev:2021jsz}. Knowledge gained during these investigations is very useful
to examine the resonances $T_{cs0}^{a0/++}$.

In our view, the resonance $T_{cs0}^{a++}$ is a more interesting object for
exploration, because it has additional attractive feature as a first doubly
charged tetraquark observed in the experiment. As is seen, the $%
T_{cs0}^{a++} $ has a content which is identical to one of the tetraquarks $%
Z^{++}$. The scalar tetraquark $Z_{\mathrm{S}}^{++}$ has the mass and width
\cite{Agaev:2017oay}
\begin{equation}
m_{\mathrm{Z}_{\mathrm{S}}}=2628_{-153}^{+166}~\mathrm{MeV},\ \Gamma _{%
\mathrm{Z}_{\mathrm{S}}}=(66.89\pm 15.11)~\mathrm{MeV,}  \label{eq:Theor1}
\end{equation}%
which are far from the parameters of $T_{cs0}^{a++}$. \

As we have noted above, the neutral scalar resonance $T_{cs0}^{a0}$ is
composed of quarks $cd\overline{s}\overline{u}$. The quark content and
spin-parity of $T_{cs0}^{a0}$ coincide with parameters of the scalar
tetraquark $\overline{X}_{c}=[cd][\overline{s}\overline{u}]$. The latter is
an antiparticle of $X_{c}$ and should have the same parameters as $X_{c}$%
\begin{equation}
m_{\mathrm{S}}=(2634\pm 62)~\mathrm{MeV}~\mbox{and}~\ \Gamma _{\mathrm{S}%
}=(57.7\pm 11.6)~\mathrm{MeV},  \label{eq:Theor2}
\end{equation}%
and
\begin{equation}
m_{\mathrm{A}}=(2590\pm 60)~\mathrm{MeV}~\mbox{and}~\ \Gamma _{\mathrm{A}%
}=(63.4\pm 14.2)~\mathrm{MeV},  \label{eq:Theor3}
\end{equation}%
which were found using for $X_{c}$ scalar-scalar or axial-axial currents .
Let us note that the prediction $(2.55\pm 0.09)~\mathrm{GeV}$ for the mass
of the $X_{c}$ was obtained in Ref.\ \cite{Chen:2016mqt}, as well. In other
words, neither $T_{cs0}^{a++}$ nor $T_{cs0}^{a0}$ can be interpreted as the
ground-level scalar diquark-antidiquark systems.

In this situation, we can study the $T_{cs0}^{a++}$ and $T_{cs0}^{a0}$
within a hadronic molecule model, i.e., as a bound state of conventional
mesons. Let us analyze in details the resonance $T_{cs0}^{a++}$.
Interpretation of a molecule $D_{s}^{+}\pi ^{+}$ as the resonance $%
T_{cs0}^{a++}$ seems is difficult, because the mass of such system is
considerably smaller than $m_{2\mathrm{exp}}$. The hadronic molecules $%
D_{s}^{\ast +}\rho ^{+}$ and $D^{\ast +}K^{\ast +}$ or their superposition
provide alternative choices for $T_{cs0}^{a++}$. Two-particle thresholds for
these molecules are equal to $2887~\mathrm{MeV}$ and $2902\ \mathrm{MeV}$,
respectively. They cannot decay to $D_{s}^{\ast +}+\rho ^{+}$ and $D^{\ast
+}+K^{\ast +}$ meson pairs if their masses are less than these thresholds.
Otherwise, the molecules $D_{s}^{\ast +}\rho ^{+}$ and $D^{\ast +}K^{\ast +}$
with masses above these limits dissociate to these mesons. In both cases,
decay to a pair of the pseudoscalar $D_{s}^{+}$ and $\pi ^{+}$ mesons are
kinematically allowed channels for the molecules $D_{s}^{\ast +}\rho ^{+}$
and $D^{\ast +}K^{\ast +}$.

In our paper \cite{Agaev:2022eyk}, we investigated the resonance $%
T_{cs0}^{a++}$ by modeling it as a the hadronic molecule $D^{\ast +}K^{\ast
+}$. The results for the mass and width of such compound $\ (2924\pm 107)~%
\mathrm{MeV},~$and $(123\pm 25)~\mathrm{MeV}$ are consistent with parameters
of $T_{cs0}^{a++}$ given by Eq.\ (\ref{eq:DtaT++}). The resonances $%
T_{cs0}^{a0/++}$ were investigated using different models and approaches in
Refs.\ \cite{Chen:2022svh,Ge:2022dsp,Wei:2022wtr,Liu:2022hbk}. It is
interesting that conclusions made about nature of these structures also
differ from each other. Thus, in Ref.\ \cite{Chen:2022svh} the one-boson
exchange model was used to explore the interactions in $D^{(\ast )}K^{(\ast
)}$ systems. Analysis performed in this article allowed the authors to
assign $T_{cs0}^{a++}$ to be an isovector $D^{\ast +}K^{\ast +}$ molecule
state with the spin-parity $J^{\mathrm{P}}=0^{+}$ and mass $2891~\mathrm{MeV}
$. Interpretation of the new tetraquark candidate $T_{cs0}^{a}$ as the
resonance-like structure induced by threshold effects was suggested in Ref.\
\cite{Ge:2022dsp}. Here, it was argued that the triangle singularity
generated by the $\chi _{c1}K^{\ast }D^{\ast }$ loop peaks around the
threshold $D^{\ast }K^{\ast }$ and may simulate $T_{cs0}^{a}$.

A multiquark color flux-tube model was used to investigate the resonances $%
T_{cs0}^{a0/++}$ in the framework of the diqark-antidiquark model \cite%
{Wei:2022wtr}. The authors found that a system $[cu][\overline{s}\overline{d}%
]$ built of the color antitriplet diquark and triplet antidiquark with the
mass $2923~\mathrm{MeV}$ is very nice candidate to the resonance $%
T_{cs0}^{a++}$. The properties of the charmed-strange tetraquarks were
studied in Ref.\ \cite{Liu:2022hbk} by employing a nonrelativistic potential
quark model.

In present work, we explore the spectroscopic parameters of the molecule $%
\mathcal{M}^{++}=D_{s}^{\ast +}\rho ^{+}$. The molecular structure $\mathcal{%
M}^{0}=D_{s}^{\ast +}\rho ^{-}$ may be considered as a model for $%
T_{cs0}^{a0}$. We calculate the mass and current coupling of $\mathcal{M}%
^{++}$ using QCD two-point sum rule method, and confront our predictions
with the experimental data of the LHCb Collaboration.

This article is structured in the following manner: In Sec.\ \ref%
{sec:MassCoupl}, we derive the sum rules for the mass $m$ and current
coupling $f$ of the molecule $\mathcal{M}^{++}$ in the context of QCD sum
rule method. Numerical analysis of the quantities $m$ and $f$ is performed
in Sec.\ \ref{sec:NAnalysis}, where we determine working windows for the
Borel and continuum subtraction parameters, and evaluate $m$ and $f$. The
section \ref{sec:Discussion} contains our concluding remarks.


\section{Spectroscopic parameters of the hadronic molecule $\mathcal{M}%
^{++}=D_{s}^{\ast +}\protect\rho ^{+}$}

\label{sec:MassCoupl}

We compute the mass $m$ and current coupling $f$ of the hadronic molecule $%
\mathcal{M}^{++}$ using the QCD two-point sum rule method \cite%
{Shifman:1978bx,Shifman:1978by}. To obtain sum rules for $m$ and $f$, we
start to the analysis by considering the following correlation function:
\begin{equation}
\Pi (p)=i\int d^{4}xe^{ipx}\langle 0|\mathcal{T}\{J(x)J^{\dag
}(0)\}|0\rangle ,  \label{eq:CF1}
\end{equation}%
where $\mathcal{T}$ is the time-ordering operator, and $J(x)$ stands for the
interpolating current of the molecule $\mathcal{M}^{++}$.

In the molecule model colorless four-quark structures come from the
singlet-singlet $\mathbf{[1}_{c}\mathbf{]\otimes \lbrack 1}_{c}\mathbf{]}$
and octet-octet $\mathbf{[8}_{c}\mathbf{]\otimes \lbrack 8}_{c}\mathbf{]}$
terms of the color group $SU_{c}(3)$. In the case of $\mathcal{M}^{++}$, we
suppose that the hadronic molecule $\mathcal{M}^{++}$ is made of two vector
mesons $D_{s}^{\ast +}$ and $\rho ^{+}$, and consider the singlet-singlet
type current. Then, in the $\mathbf{[1}_{c}\mathbf{]}_{\overline{s}c}\mathbf{%
\otimes \lbrack 1}_{c}\mathbf{]}_{\overline{d}u}$ representation $J(x)$ has
the following form
\begin{equation}
J(x)=[\overline{s}_{a}(x)\gamma ^{\mu }c_{a}(x)][\overline{d}_{b}(x)\gamma
_{\mu }u_{b}(x)],  \label{eq:CR1}
\end{equation}%
with $a$ and $b$ being color indices.

It is worth to note that $J(x)$ couples not only to the molecule $\mathcal{M}%
^{++}$ but also to diquark-antidiquark states. The reason is that a molecule
current by means of Fierz transformation can be expressed as the sum of
different diquark-antidiquark currents with some numerical factors \cite%
{Wang:2020rcx}. In other words, the molecule current is a special weighted
sum of diquark-antidiquark currents. Contrary, a diquark-antidiquark current
can be rewritten via molecule structures Refs.\ \cite%
{Chen:2022sbf,Xin:2021wcr}. Only comparison with experimental data can
justify a choice of molecule or diquark-antidiquark type structures to model
the resonance $\mathcal{M}^{++}$.

In the sum rule approach, the correlator $\Pi (p)$ has to be presented using
physical parameters of $\mathcal{M}^{++}$, and also written down in terms of
different  quark-gluon condensates of QCD. For the physical side of the sum
rule, we get
\begin{equation}
\Pi ^{\mathrm{Phys}}(p)=\frac{\langle 0|J|\mathcal{M}^{++}(p)\rangle \langle
\mathcal{M}^{++}(p)|J^{\dagger }|0\rangle }{m^{2}-p^{2}}+\cdots ,
\label{eq:Phys1}
\end{equation}%
where $p$ is the four momentum of $\mathcal{M}^{++}$. In Eq.\ (\ref{eq:Phys1}%
) the term shown explicitly is contribution of ground-state particle $%
\mathcal{M}^{++}$, whereas ellipses stand for effects of higher resonances
and continuum states in the $\mathcal{M}^{++}$ channel. To derive the
physical side $\Pi ^{\mathrm{Phys}}(p)$ of the sum rule from Eq.\ (\ref%
{eq:CF1}), we insert a complete set of intermediate states with content and
quantum numbers of the $\mathcal{M}^{++}$ state, and carry out integration
over $x$.

Introducing the physical parameters of $\mathcal{M}^{++}$ through the matrix
element
\begin{equation}
\langle 0|J|\mathcal{M}^{++}\rangle =fm,  \label{eq:ME1}
\end{equation}%
we recast $\Pi ^{\mathrm{Phys}}(p)$ into the final form%
\begin{equation}
\Pi ^{\mathrm{Phys}}(p)=\frac{f^{2}m^{2}}{m^{2}-p^{2}}+\cdots .
\label{eq:Phen2}
\end{equation}%
The function $\Pi ^{\mathrm{Phys}}(p)$ includes only one Lorentz structure,
namely the unit matrix $\mathrm{I}$, and term in rhs of Eq.\ (\ref{eq:Phen2}%
) is the invariant amplitude $\Pi ^{\mathrm{Phys}}(p^{2})$ corresponding to
this structure.

The QCD side of the sum rules, $\Pi ^{\mathrm{OPE}}(p)$, should be
calculated in the operator product expansion ($\mathrm{OPE}$) with certain
accuracy. To find $\Pi ^{\mathrm{OPE}}(p)$, we use in Eq.\ (\ref{eq:CF1})
the interpolating current $J(x)$, and contract the corresponding heavy and
light quark fields using the Wick's theorem. By performing these operations
for $\Pi ^{\mathrm{OPE}}(p)$, we obtain
\begin{eqnarray}
&&\Pi ^{\mathrm{OPE}}(p)=i\int d^{4}xe^{ipx}\mathrm{Tr}\left[ \gamma _{\mu
}S_{c}^{aa^{\prime }}(x)\gamma _{\nu }S_{s}^{a^{\prime }a}(-x)\right]  \notag
\\
&&\times \mathrm{Tr}\left[ \gamma ^{\mu }S_{u}^{bb^{\prime }}(x)\gamma ^{\nu
}S_{d}^{b^{\prime }b}(-x)\right] ,  \label{eq:QCD1}
\end{eqnarray}%
with $S_{c}(x)$ and $S_{u(s,d)}(x)$ being the quark propagators. The
explicit expressions for the heavy and light quarks propagators are
collected in Ref.\ \cite{Agaev:2020zad}.

The $\Pi ^{\mathrm{OPE}}(p)$ has also a trivial structure $\sim \mathrm{I}$
and is characterized by an amplitude $\Pi ^{\mathrm{OPE}}(p^{2})$. After
equating the invariant amplitudes $\Pi ^{\mathrm{Phys}}(p^{2})$ and $\Pi ^{%
\mathrm{OPE}}(p^{2})$, one obtains the QCD sum rule equality. In order to
suppress contributions of higher resonances and continuum states one has to
apply the Borel transformation to both sides of this expression. We apply
the continuum subtraction supported by the quark-hadron duality assumption
as well. These operations generate dependence of the sum rule equality on
the Borel $M^{2}$ and continuum threshold $s_{0}$ parameters. Final
expression and its derivative over $d/d(-1/M^{2})$ can be used to derive sum
rules for the mass $m$ and coupling $f$ of the molecule $\mathcal{M}^{++}$
\begin{equation}
m^{2}=\frac{\Pi ^{\prime }(M^{2},s_{0})}{\Pi (M^{2},s_{0})},  \label{eq:Mass}
\end{equation}%
and
\begin{equation}
f^{2}=\frac{e^{m^{2}/M^{2}}}{m^{2}}\Pi (M^{2},s_{0}).  \label{eq:Coupl}
\end{equation}%
Here, $\Pi (M^{2},s_{0})$ is the invariant amplitude $\Pi ^{\mathrm{OPE}%
}(p^{2})$ after Borel transformation and subtraction procedures, and $\Pi
^{\prime }(M^{2},s_{0})=d\Pi (M^{2},s_{0})/d(-1/M^{2})$.

The Borel transformation of the amplitude $\Pi ^{\mathrm{Phys}}(p^{2})$ has
the simple form
\begin{equation}
\mathcal{B}\Pi ^{\mathrm{Phys}}(p^{2})=fme^{-m^{2}/M^{2}},
\end{equation}%
whereas the correlator $\Pi (M^{2},s_{0})$ is given by the expression%
\begin{equation}
\Pi (M^{2},s_{0})=\int_{(m_{c}+m_{s})^{2}}^{s_{0}}ds\rho ^{\mathrm{OPE}%
}(s)e^{-s/M^{2}}+\Pi (M^{2}).  \label{eq:InvAmp}
\end{equation}%
The spectral density $\rho ^{\mathrm{OPE}}(s)$ is calculated as an imaginary
part of the amplitude $\Pi ^{\mathrm{OPE}}(p^{2})$. The term $\Pi (M^{2})$
in Eq.\ (\ref{eq:InvAmp}) is the Borel transformations of some of terms
evaluated directly from their expressions in $\Pi ^{\mathrm{OPE}}(p)$. In
this article, we neglect the masses of the $u$ and $d$ quarks, but take into
account terms $\sim m_{s}$ setting, at the same time, $m_{s}^{2}=0$.
Computations are performed by including into analysis the vacuum expectation
values of the nonperturbative operators up to dimension $8$. The higher
dimensional contributions to $\Pi (M^{2},s_{0})$ are obtained as products of
basic vacuum condensates using the factorization procedure. First few terms
in $\mathrm{OPE}$ do not contains such condensates. They appear at higher
dimensions and are numerically small. Therefore, we neglect the ambiguities
caused by the factorization and their impact on the sum rules' results.

Analytical expressions of $\rho ^{\mathrm{OPE}}(s)$ and $\Pi (M^{2})$ are
rather lengthy and not presented here explicitly.


\section{Numerical analysis}

\label{sec:NAnalysis}

The sum rules in Eqs.\ (\ref{eq:Mass}) and (\ref{eq:Coupl}) contain
different quark, gluon and mixed condensates. They are universal parameters,
and were extracted from the analysis of numerous processes. The mass $m$ and
coupling $f$ depend also on the masses of $c$ and $s$ quarks. To carry out
numerical computations, one has to fix values all of these parameters.
Below, we list the values of these condensates
\begin{eqnarray}
&&\langle \overline{q}q\rangle =-(0.24\pm 0.01)^{3}~\mathrm{GeV}^{3},\
\langle \overline{s}s\rangle =(0.8\pm 0.1)\langle \overline{q}q\rangle ,
\notag \\
&&\langle \overline{q}g_{s}\sigma Gq\rangle =m_{0}^{2}\langle \overline{q}%
q\rangle ,\ \langle \overline{s}g_{s}\sigma Gs\rangle =m_{0}^{2}\langle
\overline{s}s\rangle ,  \notag \\
&&m_{0}^{2}=(0.8\pm 0.2)~\mathrm{GeV}^{2},  \notag \\
&&\langle \frac{\alpha _{s}G^{2}}{\pi }\rangle =(0.012\pm 0.004)~\mathrm{GeV}%
^{4},  \notag \\
&&\langle g_{s}^{3}G^{3}\rangle =(0.57\pm 0.29)~\mathrm{GeV}^{6},  \notag \\
&&m_{s}=93_{-5}^{+11}~\mathrm{MeV},\ m_{c}=1.27\pm 0.02~\mathrm{GeV}.
\label{eq:Parameters}
\end{eqnarray}

\begin{figure}[h]
\includegraphics[width=8.5cm]{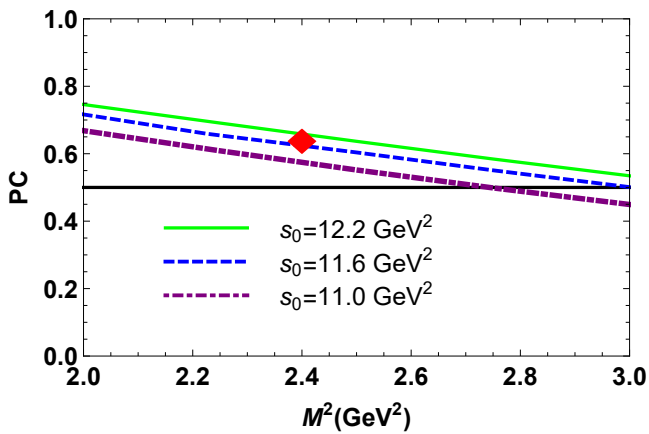}
\caption{The pole contribution to the correlator $\Pi (M^{2},s_{0})$ as a
function of the Borel parameter $M^{2}$ at different $s_{0}$. The horizontal
black line limits the border $\mathrm{PC}=0.5$. The red diamond marks the
point at which the mass $m$ of the molecule $\mathcal{M}^{++}=D_{s}^{\ast +}%
\protect\rho ^{+}$ has effectively been extracted. }
\label{fig:PC}
\end{figure}
We included into Eq.\ (\ref{eq:Parameters}) the masses of $c$ and $s$
quarks, as well.

The working regions for the parameters $M^{2}$ and $s_{0}$ should meet the
standard constraints imposed on the pole contribution ($\mathrm{PC}$) and
convergence of the operator product expansion. To quantify these
restrictions, it is convenient to introduce the formulas
\begin{equation}
\mathrm{PC}=\frac{\Pi (M^{2},s_{0})}{\Pi (M^{2},\infty )},  \label{eqPC}
\end{equation}%
and
\begin{equation}
R(M^{2})=\frac{\Pi ^{\mathrm{Dim8}}(M^{2},s_{0})}{\Pi (M^{2},s_{0})}.
\label{eq:Convergence}
\end{equation}%
First of them is a measure of the pole contribution and is necessary to find
the upper limit $M_{\mathrm{max}}^{2}$ of the Borel region. In sum rule
analyses of the conventional hadrons a constraint $\mathrm{PC}\geq 0.5$ is a
standard requirement. In the case of multiquark hadrons this constraint may
be satisfied, although it shrinks a region for $M^{2}$. The function $%
R(M^{2})$ is employed to find the lower limit, $M_{\mathrm{min}}^{2}$, of
the window for the Borel parameter. Here, $\Pi ^{\mathrm{Dim8}}(M^{2},s_{0})$
indicates the last term in the operator product expansion of $\ \Pi
(M^{2},s_{0})$. For computations performed up to dimension $8$ the
constraint $R(M_{\mathrm{min}}^{2})\leq 0.05$ is a reasonable requirement.

Analysis demonstrates that the regions for the parameters $M^{2}$ and $s_{0}$
\begin{equation}
M^{2}\in \lbrack 2,3]~\mathrm{GeV}^{2},\ s_{0}\in \lbrack 11,12.2]~\mathrm{%
GeV}^{2},  \label{eq:Wind1a}
\end{equation}%
obey all the necessary constraints. Indeed, at $M^{2}=3~\mathrm{GeV}^{2}$
the pole contribution on average in $s_{0}$ is $0.50$, whereas at $M^{2}=2~%
\mathrm{GeV}^{2}$ it becomes equal to $0.72$. In Fig.\ \ref{fig:PC} the pole
contribution is depicted as a function of $M^{2}$ at various fixed $s_{0}$.
Only at $s_{0}=11~\mathrm{GeV}^{2}$ in a small region, $M^{2}\geq 2.8~%
\mathrm{GeV}^{2}$, the pole contribution is less than $0.5$. But on average
in $s_{0}$ the constraint $\mathrm{PC}\geq 0.5$ is satisfied in the entire
working region for the Borel parameter. At the minimum point, $M^{2}=2~%
\mathrm{GeV}^{2}$, we get $R(2~\mathrm{GeV}^{2})\approx 0.022$ and the
contribution of dimension-$8$ term does not exceed $2.2\%$ of the whole
result.

Another important question which should be taken into account in fixing the
regions in Eq.\ (\ref{eq:Wind1a}) is the prevalence of the perturbative
contribution to $\Pi (M^{2},s_{0})$ over the nonperturbative one, as well as
the convergence of the operator product expansion in these regions. From
Fig.\ \ref{fig:Conv} it becomes clear that even at $M^{2}=2~\mathrm{GeV}^{2}$
the perturbative contribution constitutes more than $65\%$ of $\Pi
(M^{2},s_{0})$, whereas the nonperturbative contribution is less than $35\%$
and gradually falls with increasing of $M^{2}$. The convergence of $\mathrm{%
OPE}$ implies reducing contributions of the nonperturbative terms by
increasing dimensions of the corresponding operators. Here, one should take
into account that due to smallness of gluon condensates this hierarchy may
be destroyed for few terms. In the right panel of Fig.\ \ref{fig:Conv}, we
plot different nonperturbative contributions to $\Pi (M^{2},s_{0})$. The
contributions of terms $\mathrm{Dim3}$ and $\mathrm{Dim6}$ are positive:
contributions of other operators are negative and their absolute values do
not exceed the positive terms. The last two terms in this expansion at $%
M^{2}=2~\mathrm{GeV}^{2}$ form only $0.22$ and $0.07$ parts of $\mathrm{Dim3}
$ contribution, respectively. In other words, the operator product expansion
converges quite well.

Results for $m$ and $f$ are obtained by taking their mean values at
different choices of the parameters $M^{2}$ and $s_{0}$
\begin{eqnarray}
m &=&(2917~\pm 135)~\mathrm{MeV},  \notag \\
f &=&(4.65\pm 0.95)\times 10^{-3}~\mathrm{GeV}^{4}.  \label{eq:Result1}
\end{eqnarray}%
The $m$ and $f$ \ from Eq.\ (\ref{eq:Result1}) effectively correspond to sum
rules' predictions at $M^{2}=2.4~\mathrm{GeV}^{2}$ and $s_{0}=11.8~\mathrm{%
GeV}^{2}$ marked in Fig.\ \ref{fig:PC} by the red diamond, where the pole
contribution is $\mathrm{PC}\approx 0.64$. This fact ensures the
ground-state nature of $\mathcal{M}^{++}$  and reliability of obtained results.

The mass $m$ as functions of the parameters $M^{2}$ and $s_{0}$ is shown in
Fig.\ \ref{fig:Mass}. Here, one can see a dependence of $m$ on the parameter
$M^{2}$, though the physical quantity should not depend on it. Nevertheless,
such residual dependence of $m$ and $f$ on the Borel parameter exists and
generates essential part of theoretical uncertainties shown in Eq.\ (\ref%
{eq:Result1}). There is also dependence on the choice of $s_{0}$, but this
effect may be used to extract information on the mass of the first excited
particle in the $D_{s}^{\ast +}\rho ^{+}$ channel. Because $\sqrt{s_{0}}$
should be less than the mass $m^{\ast }$ of the first excited molecule $%
\mathcal{M}^{++\ast }$, we find an estimate $m^{\ast }\geq m+500~\mathrm{MeV}
$, which may be expected for $\mathcal{M}^{++}$ molecule containing a $c$
quark.

\begin{widetext}

\begin{figure}[h!]
\begin{center}
\includegraphics[totalheight=6cm,width=8cm]{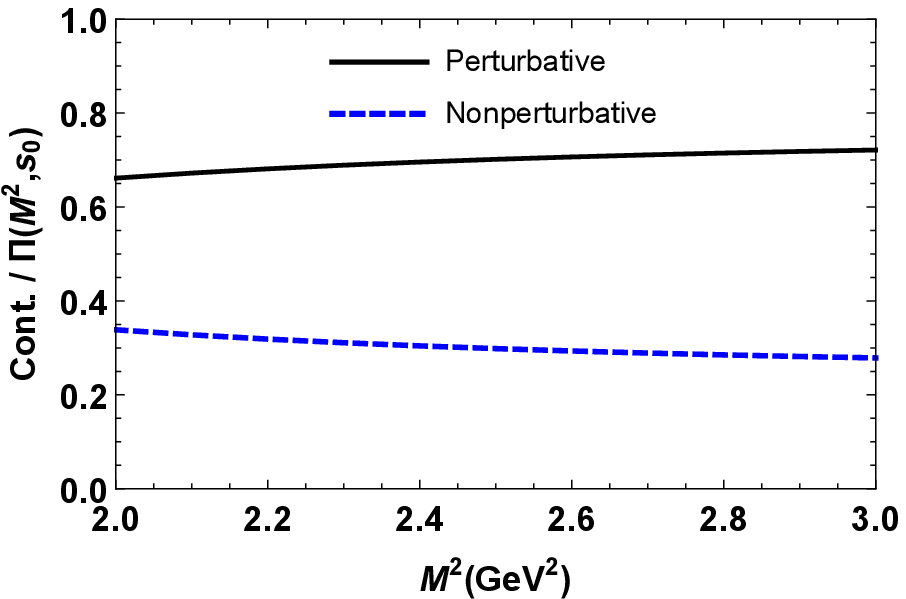}
\includegraphics[totalheight=6cm,width=8cm]{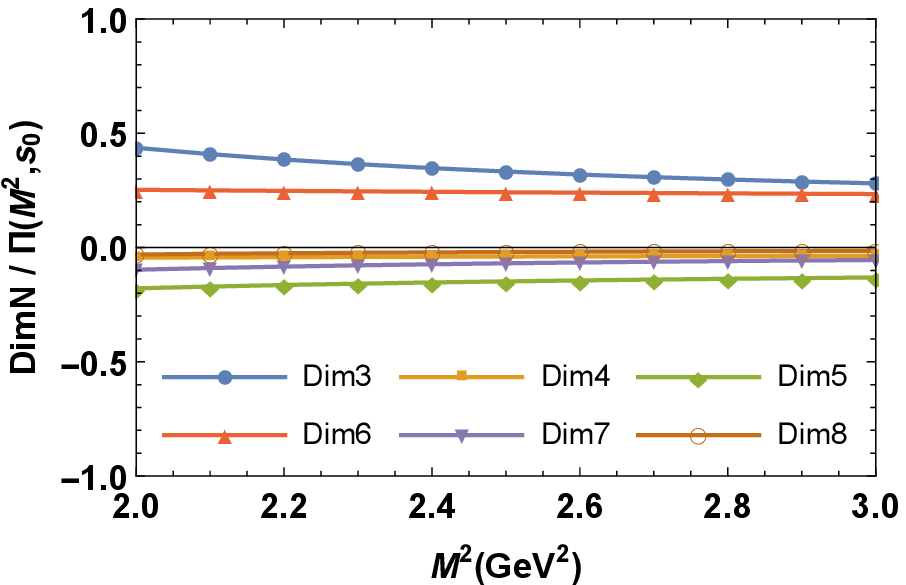}
\end{center}
\caption{Left: Perturbative and nonperturbative contributions to $\Pi
(M^{2},s_{0})$ normalized to $1$ as functions of the Borel parameter $M^2$,
Right: Normalized contributions of different operators to the correlator as functions of $M^2$. All curves in the figure have been calculated at $s_0=11.6~\mathrm{MeV}^2$.}
\label{fig:Conv}
\end{figure}

\begin{figure}[h!]
\begin{center}
\includegraphics[totalheight=6cm,width=8cm]{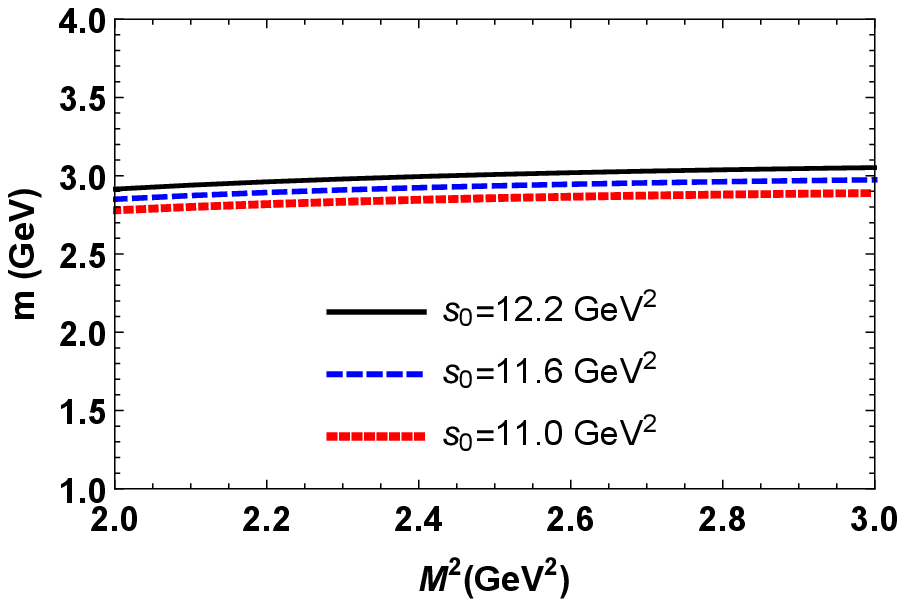}
\includegraphics[totalheight=6cm,width=8cm]{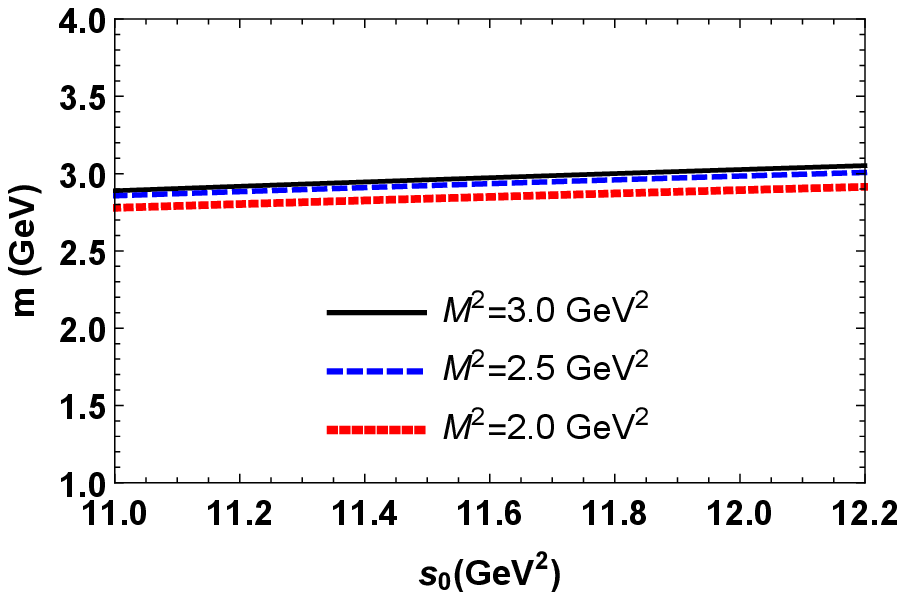}
\end{center}
\caption{The dependence of the mass $m$ of the molecule $\mathcal{M}^{++}$ on the Borel parameter $M^{2}$ (left panel), and on the continuum threshold parameter $s_0$ (right panel).}
\label{fig:Mass}
\end{figure}

\end{widetext}

\section{Concluding remarks}

\label{sec:Discussion}

As is seen, the result $m=(2917~\pm 135)~\mathrm{MeV}$ for the mass of the
molecule $\mathcal{M}^{++}=D_{s}^{\ast +}\rho ^{+}$ obtained in the current
paper agrees nicely with the LHCb datum $m_{2\mathrm{exp}}$. In our article
\cite{Agaev:2022eyk}, we modeled the resonance $T_{cs0}^{a++}$ as the
hadronic molecule $D^{\ast +}K^{\ast +}$, and calculated its mass and width.
The mass $(2924\pm 107)~\mathrm{MeV}$ of $D^{\ast +}K^{\ast +}$ is also
consistent with the LHCb data. Comparing these two models with each other,
one sees that they lead to very close results and can be applied to describe
the resonance $T_{cs0}^{a++}$. Therefore, at this level of our knowledge, we
can interpret the molecules $D_{s}^{\ast +}\rho ^{+}$ and $D^{\ast +}K^{\ast
+}$, or their superposition as candidates to the resonance $T_{cs0}^{a++}$.

The neutral molecules $\mathcal{M}^{0}=D_{s}^{\ast +}\rho ^{-}$ and $D^{\ast
0}K^{\ast 0}$ are possible models for the resonance $T_{cs0}^{a0}$. A linear
superposition of these molecules also may be used to model $T_{cs0}^{a0}$.
The mass of $T_{cs0}^{a0}$ as an isoscalar partner of $T_{cs0}^{a++}$ should
be close to the extracted value $m$. In fact, experimentally measured mass
and width differences between $T_{cs0}^{a++}$ and $T_{cs0}^{a0}$ are equal
to $\Delta m\approx 28~\mathrm{MeV}$ and $\Delta \Gamma \approx 15~\mathrm{%
MeV}$, respectively. But, to be accepted as a reliable model for $%
T_{cs0}^{a0}$ the molecule picture should be successfully confronted with
the LHCb data.

As we have noted in Sec.\ \ref{sec:MassCoupl}, the $J(x)$ couples to various
diquark-antidiquark states, and their specially chosen superposition gives
the molecule current. The fixed ground-level diquark-antidiquark state, as
it has been discussed in a detailed form in Sec.\ \ref{sec:Intro}, does not
describe parameters of the resonances $T_{cs0}^{a0/++}$. In these
circumstances, assumptions about molecule nature of $T_{cs0}^{a0/++}$ seem
are more realistic than other four-quark models. Number of alternative
molecule models for $T_{cs0}^{a0/++}$ is restricted by the masses and widths
of these states, as well as by the fact that the resonance $T_{cs0}^{a++}$
carries two units of electric charge. This question requires additional
detailed analysis, which is beyond the scope of the present work.

The structures $T_{cs0}^{a0/++}$ were studied in the context of other models
as well \ \cite{Chen:2022svh,Ge:2022dsp,Wei:2022wtr,Liu:2022hbk}. At present
there are no definite conclusion on the quark structure not only of
resonances $T_{cs0}^{a0/++}$ but also other exotic hadrons. This paper and
Ref. \cite{Agaev:2022eyk} are attemps to clarify a situation around of very
interesting structure $T_{cs0}^{a++}$.

\end{document}